\documentclass[conference]{article}
\usepackage{amssymb,amsmath}
\usepackage{epstopdf}

\usepackage{graphicx}
\graphicspath{%
    {converted_graphics/}% inserted by PCTeX
    {/}% inserted by PCTeX
}

\begin{document}

\title{{{\bf Entropy Control Architectures for\\
             Next-Generation Supercomputers}} }       
\author{{Jeffrey Uhlmann}\footnote{Dept. of EECS, University of Missouri-Columbia. 
               Position paper for US strategic computing initiative (Aug. 2017).}}
   
\date{}          
\maketitle

Progress in high-performance computing (HPC) fundamentally
requires effective thermal dissipation. At present this challenge
is viewed in terms of two distinct architectural components:
a computational substrate and a heat-exchange substrate.
Typically these two components are viewed as 
performing conceptually different functions that are independent
up to optimization constraints applied when they are jointly 
implemented. Next-generation computing architectures, by
contrast, must be designed
from the perspective of computing as a thermodynamic
process that can be managed in such a way as to actively
control {\em when} and {\em where} heat is generated.

The fundamental connection between computation and
thermodynamics is not new. It dates back to Landaur's
principle from 1961~\cite{rolf}, which says that a computational process is
thermodynamically neutral until information is destroyed.
More specifically, what is thought of as an erasure of a
bit is actually a process by which information
is irreversibly transformed to heat.

According to Landaur's principle a fully reversible 
computational process will not produce heat because
no information is destroyed (if information were
destroyed then by definition the process could not 
be reversed). Reversible logic gates have
been developed and studied~\cite{bennett,kerntopf}, 
especially recently in 
the context of quantum computing, and 
are designed to physically store information that
would otherwise be lost. Said another way, they
store the ancillary information needed for a conventional logic 
operation to be undone/reversed. 

The limitation
of reversible computing is, of course, that the 
amount of stored information will tend to increase 
without bound during execution of any nontrivial 
algorithm. What is important to notice is that
this ancillary information can be destroyed --
i.e., converted to heat -- or not at the 
{\em discretion} of the system. Beyond simple
erase or don't-erase discretion, the ancillary
information retained to allow reversibility
can, at the discretion of the system, be 
transported like any other information
from one physical memory location to another.
In principle, therefore,  information can be scheduled for
erasure and then be transported freely to physical
locations where it can be converted to heat
(erased) upon arrival with minimal impact on the ongoing
computational process. 

To summarize, conventional computing hardware
generates heat as a continuous by-product of the operation
of logic gates during execution of a program.
The precise distribution of heat is therefore a
function of the particular algorithm and is not generally
knowable {\em a~priori}, thus the heat-dissipation
substrate must be designed under an assumption
that pernicious heat accumulation (spikes) can occur
anywhere across the physical area in which
logic operations {\em may} be performed. 
This need to uniformly maintain separate 
computation and heatsink functionality within
the same physical space is 
why heat dissipation continues to limit the
density and/or frequency of logic operations. 

An entropy-controlled computing architecture 
(ECCA) would permit the
time and location at which heat is generated
to be actively controlled during the computational
process rather than letting it be indiscriminately 
generated at gate
locations with an expectation that a separate
physical system will soak it up. In the ECCA
model the ancillary bits that are conventionally
thought of as being stored as part of the operation
of each reversible logic gate are now treated as
bits of unneeded information that must be taken
away in a manner somewhat analogous to
garbage collection. More specifically, an 
unneeded bit could be sent without erasure to a 
nearby unused memory location or it
could be sent to a location at which there is
available capacity to accommodate the heat 
resulting from its erasure.

Continuing with the analogy to garbage
collection for memory management, the ECCA
processing of a given program could progress
without heat generation until storage for
waste bits has been exhausted, at which point
the computational process could be halted
until the bits are erased {\em in batch} and the resulting 
heat is dissipated. Alternatively, waste bits
could be {\em incrementally} transported for
erasure at locations selected so that heat
is generated uniformly across a physical
area and can dissipate without localized
accumulation.

Batch-mode ECCA is potentially attractive
for space-based applications in which 
heat distribution is easier to monitor than
to control and the objective is to maximize
the average rate of computation. For example,
a satellite-based computing system could
potentially afford to accumulate waste
bits during periods of sun exposure and
free them later during periods of shade.
In applications for which a fixed rate
of computation is required, e.g., for
continuous real-time dynamical control,
incremental-mode ECCA is needed.
More generally, however, both modes
can be applied jointly: batch-mode to exploit a known
model for the availability (or absence) 
of heatsink resources and incremental-mode
to actively optimize the management of
waste bits with respect to dynamically
monitored parameters of the system
(including its environment).

The ECCA model is most applicable
to supercomputing because the rate of computation
is fundamentally limited by the speed of light, so
any physical separation of logic gates to accommodate
heat dissipation increases the time
required for information to propagate between
them. Increasing the density of gates greatly
increases the distribution volatility of heat 
generation, i.e., the magnitude and frequency
of localized heat-spikes. Current technologies
for non-specific bulk heat dissipation using
static heat sinks or dynamic fluid transport
cannot be scaled to the decreasing length-scales needed
to maintain the current rate of improvement
in supercomputing power. Eventually a coherent
solution to the compute-vs-heat problem must
be applied. 

Developing a fully-functional ECCA
solution requires physical gates capable
of creating ancillary bits like
standard reversible logic gates but 
with means for those bits to be 
accessed and transported for
heat management. Algorithms to exploit
this available means for
effectively managing the incremental and/or
batch erasure of waste bits are clearly also
needed. Fortunately, neither the hardware
nor software aspects of the proposed 
solution involve any potentially fundamental
limiting theoretical or practical obstacles, so 
the case for moving quickly toward this next
generation of supercomputing architectures 
is strong.


\begin{thebibliography}{}

\bibitem{rolf}
Rolf R. Landauer, ``Irreversibility and heat generation in the computing process,'' 
{\em IBM Journal of Research and Development}, vol. 5, pp. 183-191, 1961.

\bibitem{bennett}
C.H. Bennett, ``Logical reversibility of computation,'' 
{\em IBM Journal of Research and Development}, vol. 17, no. 6, pp. 525-532, 1973.

\bibitem{kerntopf}
P. Kerntopf, ``Synthesis of multipurpose reversible logic gates,''
{\em Proc. Euromicro Symposium on Digital System Design}, Sept. 4-6, 2002.




\end{thebibliography}
\end{document}